\documentclass[
amsmath,
amssymb,
aps, 
pra,
twocolumn, 
groupedaddress,
nobalancelastpage,
floatfix]
{revtex4-2}
\usepackage{mathtools}
\usepackage[colorlinks]{hyperref}
\usepackage{lipsum}
\usepackage{bbold}
\usepackage{bm}
\usepackage{dsfont}
\usepackage{enumerate}
\usepackage{graphicx}
\usepackage{float}
\usepackage {xcolor}
\usepackage{cleveref}
\usepackage{amsmath}

\usepackage{lipsum}

\DeclareMathOperator{\e}{e}					

\DeclarePairedDelimiterX{\mean}[1]{\langle}{\rangle}{
	{#1}
}
\DeclarePairedDelimiterX{\abs}[1]{\lvert}{\rvert}{
	{#1}
}
\DeclarePairedDelimiterX{\norm}[1]{\lVert}{\rVert}{
	{#1}
}
\DeclarePairedDelimiterX{\bra}[1]{\langle}{\rvert}{#1}

\DeclarePairedDelimiterX{\ket}[1]{\lvert}{\rangle}{#1}

\DeclarePairedDelimiterX{\mel}[3]{\langle}{\rangle}{
	{#1}\delimsize\vert {#2}\delimsize\vert {#3}
}
\DeclarePairedDelimiterX{\inner}[2]{\langle}{\rangle}{
	{#1} \delimsize\vert{#2}
}
\DeclarePairedDelimiterX{\dyad}[2]{\lvert}{\vert}{
	{#1} \delimsize\rangle \delimsize \langle{#2}
}

\begin{document}
	\title{Edge states and quantum optical high-harmonic generation from topological insulators}
	
	\author{Christian Saugbjerg Lange}			
	\author{Lars Bojer Madsen}					
	\affiliation{Department of Physics and Astronomy, Aarhus University, Ny Munkegade 120, DK-8000 Aarhus C, Denmark}
	\date{\today}
	\begin{abstract}
		The strong-field process of high-harmonic generation (HHG) has, in recent years, been treated from a quantum optical perspective in the emerging research area of strong-field quantum optics. These investigations show that HHG radiation is, in general, in a nonclassical state of light. However, the quantum optical treatment of HHG from topological nontrivial materials is missing. Here, we aim to address this gap in current knowledge and consider the quantum optical HHG response from the Su-Schrieffer-Heeger model, a finite chain of atoms with both a topologically trivial and nontrivial insulating phase, the latter supporting edge states. We find that HHG from both topological phases is squeezed at the band-gap frequency. Interestingly, while the harmonic spectrum discriminates the two topological phases of the system, the degree of squeezing only discriminates the phases for smaller chain lengths. We attribute this difference to a relative increase in overlap between bulk and edge states in the topological nontrivial phase for smaller systems. Our findings reveal how the strength of dipole couplings governs the nonclassical HHG response and define new research questions on topologically protected generation of quantum light in strong-field physics.
	\end{abstract}
	
	\maketitle
	
		In recent years, strong-field physics has experienced a renewal by adopting a full quantum treatment of both light and matter. This emerging research area of strong-field quantum optics opens up new research questions with respect to the quantum light-matter interaction in strong-field processes; a challenging domain involving a macroscopic number of photons and a large number of electronic states, including the continuum. In this regard, high-harmonic generation (HHG), a nonlinear process where an intense laser field drives an electronic medium nonperturbatively, which in turn generates light at new frequencies, has been of particular interest with the aim of describing the generated light quantum mechanically. Studies have shown that by driving the electronic medium with an intense coherent state, the HHG light is, in general, in a nonclassical state, evident from a nonzero degree of quadrature squeezing and non-Poissonian photon statistics \cite{Gorlach2020, Lange2024a, Stammer2024a, Yi2024, Stammer2025, Lange2025c}. Quantum optical HHG has been studied from both atomic ensembles \cite{Gorlach2020, Yi2024, Rivera-Dean2024e, Stammer2025, Lange2025c}, molecules \cite{Rivera-Dean2024d}, correlated materials \cite{Lange2024a, Lange2025a, Lange2025b, Lange2025c}, and entangled systems \cite{Pizzi2023} with recent experimental verification \cite{Gonoskov2016, Tsatrafyllis2017, Tsatrafyllis2019, Theidel2024}. Additionally, it has been shown that HHG is first-order coherent \cite{Stammer2025} and that the origin of the nonclassical response is due to time correlations in the electronic medium \cite{Stammer2024a, Lange2025a, Stammer2025, Lange2025c}.	In a parallel line of studies, HHG driven by nonclassical states of light has been explored, both theoretically \cite{Gorlach2023, EvenTzur2023, EvenTzur2024, Gothelf2025, Gonzalez-Monge2025, Gothelf2026} and experimentally \cite{Rasputnyi2024, Lemieux2025, EvenTzur2025}, showcasing the broad range of new research questions to investigate with a fully quantized description of intense light-matter interactions.	 
			
		A certain class of important materials has not been considered in strong-field quantum optics, namely that with a nontrivial topological phase. The topological phases of matter are characterized by distinct quantum properties, most notably symmetry-protected edge states exponentially localized at the edges or surfaces of finite systems. These edge states are robust against perturbations that do not break the relevant symmetries \cite{Hasan2010, Xiao2011}. Consequently, topological materials have attracted interest for applications such as quantum information storage \cite{Nayak2008, Aghaee2025}. Likewise, the nonclassical response in HHG from topologically nontrivial materials, and particularly the symmetry-protected edge states, is of particular interest. 
		
		In this work, we therefore investigate the quantum optical properties of light generated via HHG in finite-sized topologically trivial and nontrivial insulators. Specifically, we examine whether the nonclassical HHG response discriminates between the two topological phases and assess how the presence of edge states in the nontrivial phase influences the quantum properties of the emitted light. We find that the quantum response, exemplified by the degree of squeezing, from HHG only discriminates the topological phases of the matter system for sufficiently small system sizes, while for larger systems, the two phases yield the same degree of squeezing across the harmonic spectrum. Our findings show how the degree of squeezing in the emitted HHG is particularly sensitive to strong dipole couplings in the matter system.

	\begin{figure}
		\centering
		\includegraphics[width=1\linewidth]{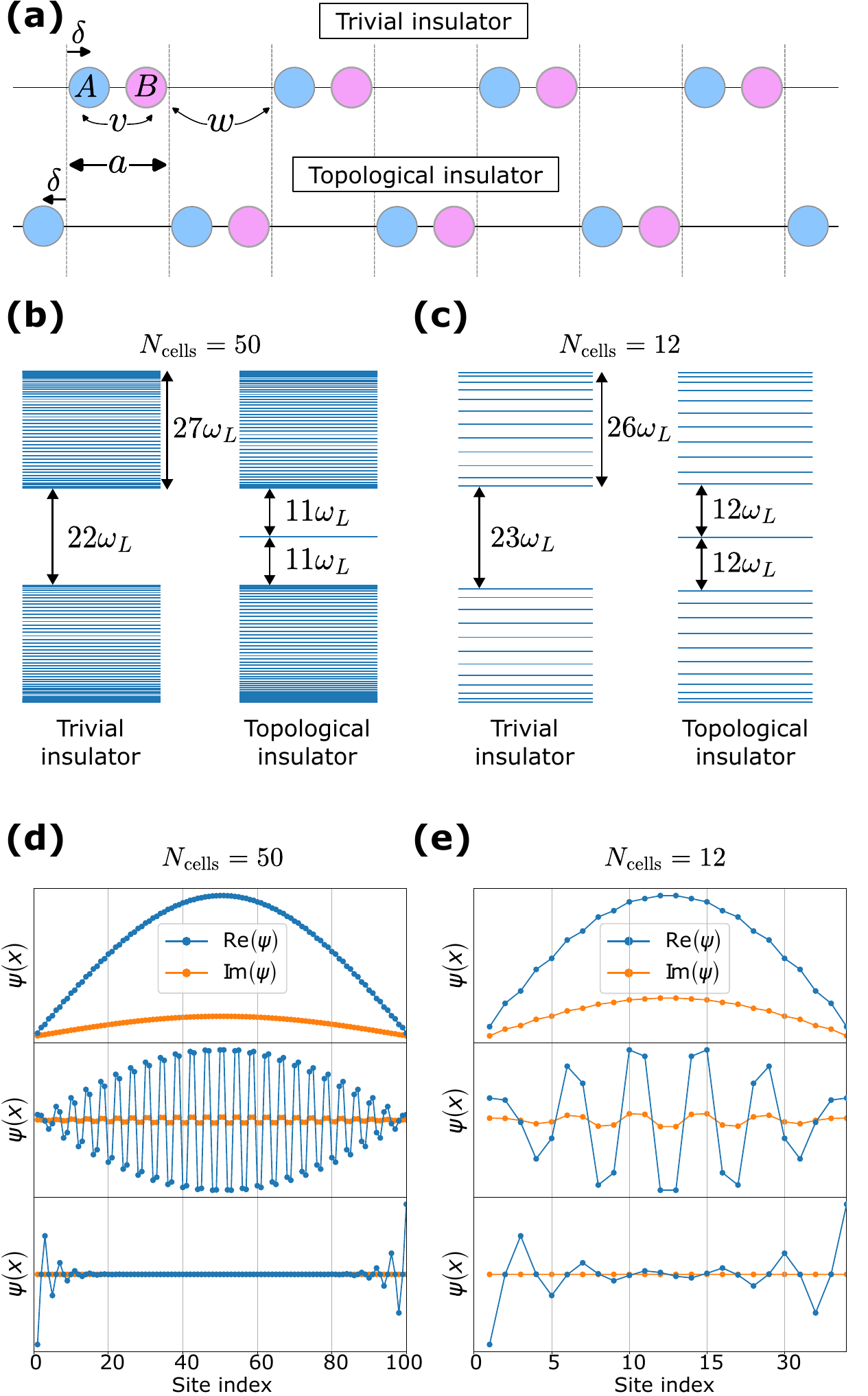}
		\caption{(a) Illustration of the real-space configuration of the SSH model in its different phases with physical interpretation of relevant parameters introduced in he main text. (b), (c) Energy spectra for the trivial and topological phases for the long (b) and short (c) SSH chains. The topological phase supports states at zero energy. (d), (e) Eigenstates of the SSH model in the topological phase for both the ground state (top row), the highest-energy state below the band gap (middle row), and the zero-energy state (bottom row), localized at the edges of the chain.}
		\label{fig:sshsystem}
	\end{figure}

We consider the Su-Schrieffer-Heeger (SSH) model, originally developed to describe long chains of polyacetylene \cite{Su1979}. We follow Refs. \cite{Bauer2018, Drueke2019, Jurss2019} and consider spinless fermions in a chain of $N_{\text{cells}}$ unit cells, each consisting of two sites, denoted $A$ and $B$. Each site is placed at $x_j = j a - (-1)^j \delta$, where $j = 1, 2, \dots, 2N_{\text{cells}}$ is the site index in the chain, $a$ the lattice constant, and $\delta$ describes an alternating shift  site location, as seen in Fig. \ref{fig:sshsystem}(a). The SSH Hamiltonian reads
	\begin{align}
		\hat{H} \! =\!\!\!  \sum^{N_{\text{cells}}}_{i=1} \!\!\! \big( v  \hat{c}_{i,A}^\dagger \hat{c}_{i,B} + \text{h.c.}\big) \!\!+ \!\!\!\!\! \sum^{N_{\text{cells}}-1}_{i=1} \!\!\!\big( w  \hat{c}_{i,B}^\dagger \hat{c}_{i+1,A} + \text{h.c.} \big), \label{eq:SSH_Hamiltonian}
	\end{align}
	where $v  = -e^{-(a-2\delta)}$ and $w  = -e^{-(a+2\delta)}$ describes the intracell and intercell hopping, respectively \cite{Jurss2019, Graf1995}, and $\hat{c}_{i,\alpha}$ ($\hat{c}_{i,\alpha}^\dagger$) is the fermionic annihilation (creation) operator for site $\alpha = A,B$ in unit cell $i$. For $\delta >0$, the system is a topologically trivial insulator, while for $\delta < 0$, the system is in a topological nontrivial phase. As polyacetylene  can be fabricated in many different lengths \cite{Lichtmann1980, Park1992, Rybachuk2008}, we consider both a relatively long ($N_{\text{cells}} = 50$) and short ($N_{\text{cells}} = 12$) SSH chain. The energy spectra for the two different topological phases are seen in Figs. \ref{fig:sshsystem} (b) and (c). The topological nontrivial phase hosts two states at $E_{\text{edge}} \approx0$, not found in the topological trivial phase. These zero-energy edge states are exponentially localized at the edge of the chain and protected by the chiral symmetry of the SSH Hamiltonian, meaning that they remain unaffected by any perturbation that does not break the chiral symmetry. In Figs. \ref{fig:sshsystem} (d) and (e), the wave functions for the ground state (top row), highest occupied bulk state at half filling (middle row), and the occupied zero-energy edge state (bottom row) are shown for the topological nontrivial phase for $N_{\text{cells}} = 50$ and $12$, respectively. The two top rows in Figs. \ref{fig:sshsystem} (d) and (e) show that the eigenstates are localized in the bulk of the system. This is also the case for all the states in the topologically trivial phase. In contrast, the bottom row shows how the zero-energy states are superposition of the two localized edge states, $(\ket{R} \pm \ket{L})/\sqrt{2}$, where $\ket{L/R}$ is the exponential localized state on the left/right edge, characteristic of the topological nontrivial phase. Though the edge states are localized for both system sizes, they extend relatively further into the bulk for $N_{\text{cells}}=12$, than for $N_{\text{cells}}=50$, see Figs. \ref{fig:sshsystem}(d) and (e).
	
	We consider the quantum optical properties of the HHG radiation from driving the SSH chain by an intense laser field. In recent work, the quantum optical response was derived using what we called the perturbative Heisenberg dynamics (PHD) formulation for strong-field quantum optics. This methodology relates the optical observables of interest to the semiclassically induced dynamics in the system \cite{Lange2025c}. This is achieved in two steps: first, the intense driving field (coherent-state parameter $\lvert \alpha_L \rvert \gg 1$) is treated exactly via a transformation resulting in semiclassical dynamics; second, the coupling to the quantized field is treated perturbatively to second order in the weak light-matter coupling constant $g_0 = 4 \times 10^{-8}$ a.u. \cite{Gorlach2020, Lewenstein2021, Lange2025a, Lange2025b, Lange2025c} . From this, it was found, confirming \cite{Sundaram1990, Stammer2025}, that the harmonic spectrum in the limit of many identical and independent emitter systems, $N$, is given by the coherent spectrum 
	
	\begin{equation}
		S_{\text{coh}}(\omega_k)  \propto N^2 \sum_\sigma \bigg \lvert \int_{-\infty}^{\infty} dt' e^{-i\omega_k t'} \langle \hat{j}_{\sigma, sc}(t')\rangle \bigg \rvert^2, \label{eq:spectrum}
	\end{equation}
	where $ \hat{j}_{\sigma, sc} = \hat{\bm{j}}_{sc} \cdot \hat{\bm{e}}_{\sigma}$ is the semiclassical current operator projected along the polarization vector $\hat{\bm{e}}_{\sigma}$ for the polarization $\sigma$. We note that Eq. (\ref{eq:spectrum}) is proportional to the semiclassical expression for the harmonic spectrum, proving that no additional quantum optical insights are provided by the spectrum \cite{Lange2025a, Lange2025c}. In contrast, a nonvanishing degree of squeezing is a clear indication of a nonclassical state \cite{Gerry2004}. We calculate the degree of squeezing in the unit of decibel via 
	\begin{equation}
		\eta_{\bm{k}, \sigma} = -10 \log_{10} \bigg \{\underset{\theta_{\bm{k}, \sigma} \in (0, \pi])}{4 \text{min}} \big[\Delta \hat{X}_{\bm{k}, \sigma}(\theta_{\bm{k}, \sigma}) \big]^2 \bigg \}, \label{eq:eta_squeezing}
	\end{equation}
	which is a minimization of the quadrature variance, which in the PHD for strong-field quantum optics is given by
	\begin{align}
		& [\Delta \hat{X}_{\bm{k}, \sigma}(\theta_{\bm{k}, \sigma})]^2  = \dfrac{1}{4} + \dfrac{g_0^2 N}{2 \omega_k} \bigg\{ \nonumber \\
		&\int_0^t dt' \int_0^t dt'' e^{-i \omega_k (t'-t'')} \langle \Delta \hat{j}_{ \sigma, sc}(t') \Delta \hat{j}_{ \sigma, sc}(t'') \rangle \nonumber \\
		& 
		- \text{Re} \bigg[  \int_0^t dt' \int_0^{t} dt'' e^{-i \omega_k (2t -t' -t'') - 2i \theta_{\bm{k}, \sigma}} \nonumber \\
		& \qquad \qquad \qquad \qquad \times \langle \Delta \hat{j}_{\sigma, sc}(t') \Delta \hat{j}_{\sigma, sc}(t'') \rangle  \nonumber\\
		& -	\int_0^t dt' \int_0^{t'} dt'' e^{-i \omega_k (2t -t' -t'') - 2i \theta_{\bm{k}, \sigma}} \nonumber \\
		& \qquad \qquad \qquad \qquad \times  \langle [\Delta \hat{j}_{\sigma, sc}(t''), \Delta \hat{j}_{\sigma, sc}(t')] \rangle
		\bigg] \bigg\}, \label{eq:quadrature_variance}
	\end{align}
	 where $\Delta \hat{j}_{\sigma, sc}(t) = \hat{j}_{\sigma, sc}(t) -\langle \hat{j}_{\sigma, sc}(t) \rangle$ expresses the fluctuations of the semiclassical current operator. If $\eta_{\bm{k}, \sigma} > 0$, the light is quadrature squeezed and nonclassical. Interestingly, the PHD predicts that the coherent spectrum in Eq. (\ref{eq:spectrum}) scales with $N^2$ and that the quadrature variance related to the squeezing scales linearly with $N$. Thus, both the signal and degree of squeezing increases with the number of emitter systems involved in the HHG process. 
	
	To evaluate Eqs. (\ref{eq:spectrum})-(\ref{eq:quadrature_variance}) in the SSH model, we drive the system by applying a classical laser field, $A_{cl}(t)$ with polarization along the SSH chain, and exploit the Peierl's phase substitution to obtain time-dependent hopping parameters $v(t) = v \e^{-i(a-2\delta)A_{cl}(t)} $ and $w(t) = w \e^{-i(a+2\delta)A_{cl}(t)}$ such that the SSH Hamiltonian in Eq. (\ref{eq:SSH_Hamiltonian}) becomes a time-dependent semiclassical Hamiltonian, $\hat{H}_{sc}(t)$ \cite{Graf1995, Jurss2019}. The current operator of the semiclassical SSH model is given as $\hat{\bm{j}}_{sc}(t) = i [\hat{H}_{sc}(t), \hat{\bm{x}}]$, where we take $\hat{\bm{x}}$ to be the direction of the chain as well as the only relevant polarization direction. We denote the single-particle eigenstates as $\ket{\psi_m} = \hat{\gamma}_{m} \ket{0}$ with $\hat{\gamma}_m = \sum_{i, \alpha} a_{i,\alpha}^{(m)} \hat{c}_{i,\alpha}^\dagger \ket{0}$ the single-particle operator of eigenstate $m$ with $a^{(m)}_{i,\alpha}$ the amplitude of the $m$'th eigenstate on site $\alpha = A,B$ in unit cell $i$. As the SSH model consists of independent particles, the many-body state is a single Slater determinant which we write as $\ket{\Psi_{M}} = \prod_{m \in M} \hat{\gamma}_{m}\ket{0}$, where $M$ denotes a many-body index consisting of a specific combination of single-body states, e.g.,  $G = \{1, 2,..., N_{\text{cells}}\}$ with $M= G$ being the many-body ground state. We only consider the system at half filling, which means that all single-particle states below the energy gap are initially occupied in the topological trivial phase, while also one of the zero-energy edge states is occupied in the topological nontrivial phase, see Figs. \ref{fig:sshsystem} (b)-(e). 
	
	To calculate the quadrature variance [Eq. (\ref{eq:quadrature_variance})], we insert a complete set of many-body eigenstates, $\mathbb{1} = \Sigma_{M} \ket{\Psi_{M}} \bra{\Psi_{M}}$, such that the expectation value of the current correlations at two different times reads \cite{Lange2025a, Lange2025c}
	
	\begin{align}
		&\langle \Delta \hat{j}_{\sigma, sc}(t') \Delta \hat{j}_{\sigma, sc}(t'') \rangle
		= \sum_{M \neq I} j_{I, M}(t')  j_{M, I} (t''), \label{eq:correlation_to_transition_currents}
	\end{align}
	where $j_{I,M}(t) = \bra{\Psi_I} \hat{\mathcal{U}}^\dagger_{sc}(t)\hat{j}_{\sigma, sc}(t) \hat{\mathcal{U}}_{sc}(t) \ket{\Psi_M}$ denotes a transition current between the many-body eigenstate $\ket{\Psi_M}$  and the initially occupied many-body state $\ket{\Psi_I}$, which we take to be the ground state in this work and $\hat{\mathcal{U}}_{sc}(t) = \otimes_i ~ \hat{\mathcal{U}}^{(1)}_{sc}(t)$ is the semiclassical time-evolution operator of the many-body state with $\hat{\mathcal{U}}^{(1)}_{sc}(t)$ being  the single-particle time-evolution operator. Due to the fact that the SSH consists of independent particles and both the Hamiltonian and current operator are single-particle operators, the SSH model is subject to Slater-Condon rules \cite{Helgaker2000molecular}. This means that only transition currents where $\ket{\Psi_M}$ differs from $\ket{\Psi_I}$ by at most a single-particle occupation need be considered. Hence, as the summation in Eq. (\ref{eq:correlation_to_transition_currents}) is for all excited many-body states $M\neq I$,  if $a \in I$ and $b \in M$ with $a \notin M$ and $b \notin I$, we find   
	\begin{equation}
		j_{I,M}(t) = \bra{\psi_a}\hat{\mathcal{U}}^{(1)\dagger}_{sc}(t) \hat{j}_{\sigma}(t) \hat{\mathcal{U}}^{(1)}_{sc}(t)\ket{\psi_b}, \label{eq:specific_transition_current}
	\end{equation}
	that is, only single-particle excitations need to be considered. To calculate the spectrum in Eq. (\ref{eq:spectrum}), $M=I$ and the expectation value of the current $\langle \hat{j}_{\sigma, sc}(t) \rangle = \bra{\Psi_I} \hat{\mathcal{U}}^\dagger_{sc}(t) \hat{j}_{\sigma, sc}(t) \hat{\mathcal{U}}_{sc}(t)  \ket{\Psi_I} = \sum_{m \in I} \bra{\psi_m} \hat{\mathcal{U}}^{(1) \dagger}_{sc}(t)\hat{j}_{\sigma, sc}(t) \hat{\mathcal{U}}^{(1)}_{sc}(t) \ket{\psi_m}$ is the sum of single-particle contributions to the many-body current. 
	
	We propagate each single-particle eigenstate according to the time-dependent Schrödinger equation using the Arnoldi-Lancoz algorithm \cite{Park1986, Smyth1998, Guan2007, Frapiccini2014} with a time step of $dt = 1.00$ a.u. and a Krylov subspace dimension of $6$. We fix the lattice spacing at $a=2.00$ a.u. and consider $\delta = \pm 0.15$ a.u., with the negative sign corresponding to the topological nontrivial phase. We drive the chain with an intense laser on the form $A(t) = (F_0/\omega_L) f(t) \cos(\omega_L t)$, where $f(t)$ is an envelope with $N_{on} = 5$ laser cycles in the $\sin^2$ ramp and $N_{pl} = 10$ laser cycles on a flat-top plateau. Similar to Refs. \cite{Bauer2018, Drueke2019, Jurss2019}, we use a laser frequency of $\omega_L= 0.0075$ a.u., corresponding to a wavelength of $\lambda_L = 6.1\mu$m. For the long chain of $N_{\text{cells}}=50$, we use an electric field amplitude of $F_0 = 0.0025$ a.u., while for the short chain of $N_{\text{cells}}=12$, an electric field amplitude of $F_0 = 0.0015$ a.u. is used. All results are checked for numerical convergence. 
	
	\begin{figure}
		\centering
		\includegraphics[width=1\columnwidth]{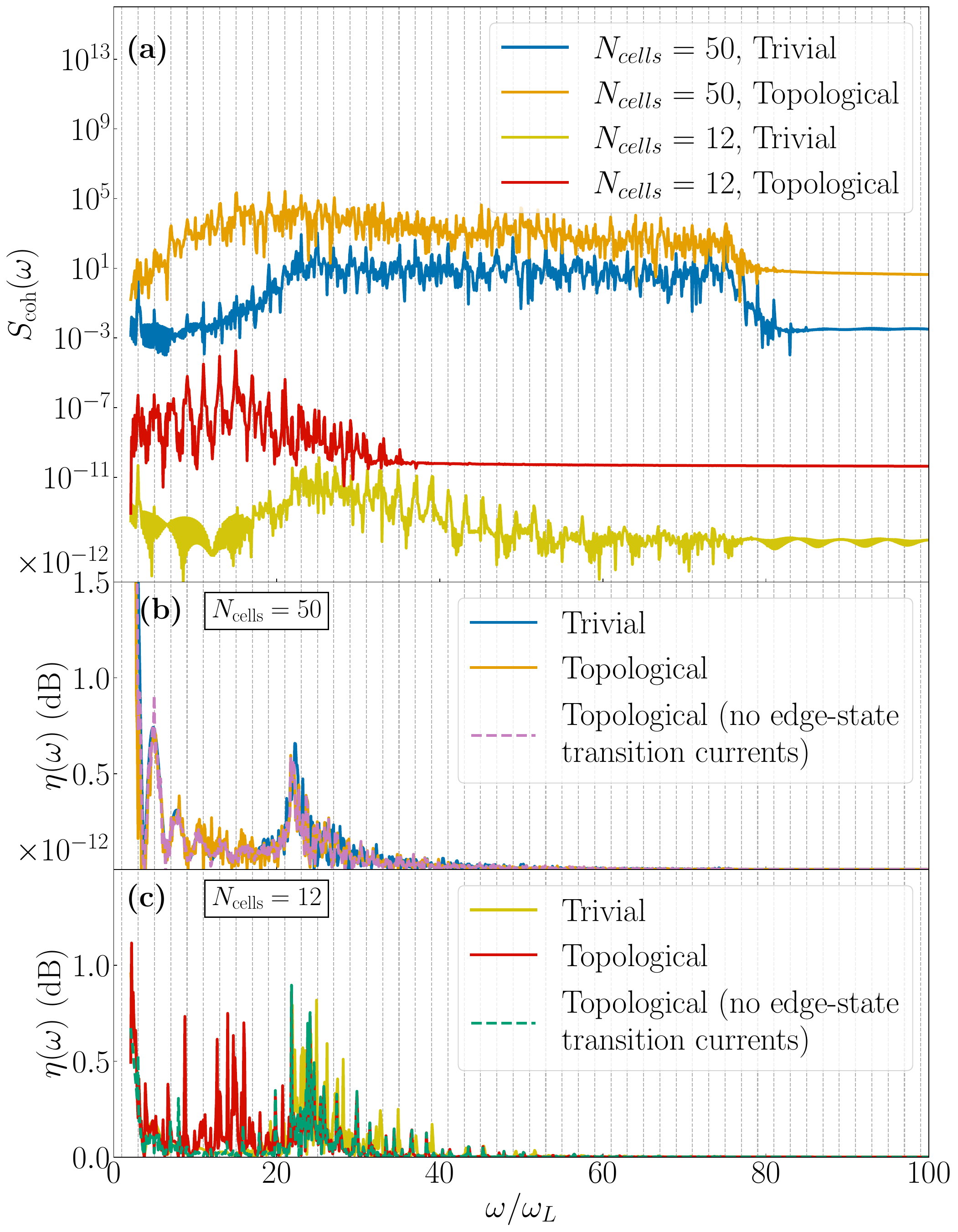}
		\caption{(a) Spectra obtained from Eq. (\ref{eq:spectrum}) from both the long (blue and orange) and short (yellow and red) chains for both the trivial ($\delta = 0.15$) and topological ($\delta = -0.15$) phase. We note how the spectra discriminate the topological phases. The degree of squeezing for the long (b) and short chain (c), respectively. We note that excluding edge-state transition currents changes the degree of squeezing for the short chain at harmonics below the gap (c, green dashed) while it leaves the degree of squeezing from the long chain (b, purple dashed) indifferent. The coherent spectra in (a) are shown without $N$ scaling and have been shifted for visual clarity, and the degree of squeezing is shown for $N=1$ SSH chain. The parameters for the system are stated in the main text.}
		\label{fig:sshspectraandsqueezing}
	\end{figure}
	
	In Fig. \ref{fig:sshspectraandsqueezing}(a), the HHG spectra are shown for the long and short chains in both topological phases. The origin of the different spectra is well understood \cite{Bauer2018, Jurss2019, Drueke2019}: the so-called interband mechanism, where a particle is promoted across the band gap, leaving a hole behind, propagates within the upper band before recombining with the hole, can only efficiently generate harmonics above the band-gap energy. This is why the topologically trivial phase has an increased signal at harmonics at around $\omega_k/ \omega_L \approx E_{gap}/\omega_L \approx 22,23$ [see Figs. \ref{fig:sshsystem} (b) and (c)] for the long and short chains, respectively. On the other hand, due to the presence of the edge states within the band gap, the interband mechanism can generate harmonic radiation down to the $11$'th and $12$'th harmonic in the topological phase both chain lengths. We note that harmonic peaks are seen below the band-gap energies for both topological phases. This is due to propagation within the band, i.e., between spectrally close-lying states above and below the energy gap [see in Figs. \ref{fig:sshsystem} (b) and (c)]. Finally, we note that for both system sizes, the topological nontrivial phase shows a stronger signal across all frequencies, which is due to transitions via the zero-energy states, resulting in a higher population of states above the gap, yielding a stronger current and hence stronger signal.
	
	The degree of squeezing [Eq. (\ref{eq:eta_squeezing})] from the two different phases is shown in Figs. \ref{fig:sshspectraandsqueezing}(b) and (c) for $N_{\text{cells}}=50$ and $12$, respectively. The degree of squeezing between the two phases is practically indifferent for $N_{\text{cells}}=50$, preventing discrimination of the topological phase of the model for this chain length, although such a discrimination is possible from the harmonic spectrum. To investigate the contribution from the edge states in the degree of squeezing, we calculate the quadrature variance [Eq. (\ref{eq:quadrature_variance})] using the transition currents in Eq. (\ref{eq:correlation_to_transition_currents}), but omitting all single-particle edge states in the summation. In other words, if either $\ket{\psi_a(t)}$ or $\ket{\psi_b(t)}$ in the resulting transition current in Eq. (\ref{eq:specific_transition_current}) is an edge state, the term is omitted in Eq. (\ref{eq:correlation_to_transition_currents}). The degree of squeezing for this case is shown by the purple dashed line in Fig. \ref{fig:sshspectraandsqueezing} (b). Omitting the edge-state transition currents does not change the degree of squeezing. This is due to the fact that the dominating transition currents in the summation in Eq. (\ref{eq:correlation_to_transition_currents}) are those between occupied bulk states below the energy gap to unoccupied bulk states above the gap. In contrast, both bulk-edge and edge-edge transition-current matrix elements are negligible in comparison. Looking at Fig. \ref{fig:sshsystem}(d), we see that this is expected, as the bulk states have very little overlap with the edge states, which remain exponentially localized on the edge of the system during the dynamics. We thus conclude that when the bulk-edge contribution in the transition currents is negligible, the degree of squeezing does not differ between the two topological phases of the SSH model. For comparison, we consider the degree of squeezing when driving a shorter chain of $N_{\text{cells}}=12$, shown in Fig. \ref{fig:sshspectraandsqueezing}(c). Here we note that the topological nontrivial phase (red) exhibits a larger degree of squeezing than the topological trivial phase (yellow) at harmonics $\omega/\omega_L \sim 10 - 20$, which is below the band-gap energy, see Fig. \ref{fig:sshsystem}. Again, to investigate the role of the edge states in this system, we calculate the degree of squeezing when removing this edge-state contribution in the transition currents, and show the results in Fig. \ref{fig:sshspectraandsqueezing}(c) (dashed green line). Here, we see that without the edge-state contributions, the degree of squeezing from the topological phase becomes identical to that found in the trivial phase, proving that the edge states are responsible for the signal at $\omega/\omega_L \sim 10 - 20$ discriminating the two phases. This is due to the fact that, although the edge states are exponentially localized in the topologically nontrivial phase, the small system size makes the overlap with bulk states non-negligible, an overlap not present in the topologically trivial system. Comparing the findings in Figs. \ref{fig:sshspectraandsqueezing}(b) and (c), we conclude that the degree of squeezing in HHG discriminates the topological phases of the SSH model only if the system size makes the bulk-edge overlap non-negligible. 
	
	Our results demonstrate that strong-field quantum optics provides direct access to nonclassical electronic responses. Within the PHD formalism for strong-field quantum optics \cite{Lange2025c}, Eq. (\ref{eq:spectrum}) shows that the emitted spectrum is proportional to the induced current, where topological edge states give a dominant contribution through an enhanced interband processes, consistent with semiclassical predictions \cite{Bauer2018, Drueke2019, Jurss2019}, and clearly distinguishes the topological phases, independent of chain length. Going beyond this classical observable, Eqs. (\ref{eq:eta_squeezing})-(\ref{eq:quadrature_variance}) show that squeezing is governed by time correlations of the induced current \cite{Stammer2024a, Lange2025a, Stammer2025, Lange2025c}. For long chains, the degree of squeezing does not discriminate the topological phases due to the vanishing bulk–edge overlap. In contrast, such a bulk-edge overlap is nonvanishing in shorter chains, allowing one to discriminate the topological phase in this latter case. Comparing the SSH model with recent work on quantum-optical HHG in excitonic systems \cite{Lange2025b} shows a stark contrast: In Ref. \cite{Lange2025b}, a strong dipole coupling between the ground state and exciton states yielded pronounced squeezing at the transition energy, while such a signature is not seen in the response from long-chain SSH model, independent of the presence of edge states. This difference originates from weaker dipole couplings in the SSH model, particularly for edge states, suppressing their contribution to the fluctuations of the induced current and hence the degree of squeezing. Consequently, we conclude that the current time correlations [Eq. (\ref{eq:correlation_to_transition_currents})] are particularly sensitive to strong dipole couplings in the matter system, and as such, the degree of squeezing in the emitted HHG probes different matter dynamics beyond those captured by the classical harmonic spectrum.
	
	Our work raises new research questions with respect to the nonclassical response in HHG from topologically nontrivial systems. Further studies include topological materials with a nonzero Chern number where the nonclassical response might yield a universal signature of the topological phase, previously not found \cite{Neufeld2023}. Further, since the edge states are symmetry-protected in the topological nontrivial phase, our work is a first step into an exploration of topologically protected quantum light via HHG, which could provide a stable source of quantum light at high frequencies with applications in, e.g., quantum sensing and technology.  
	
	This work is supported by the Novo Nordisk Foundation Project Grants in the Natural and Technical Sciences (0094623) and the Independent Research Fund Denmark (Technology and Production Sciences 10.46540/4286-00053B)

	\bibliography{big_bibliography}

\end{document}